# Ultrasound sensing at thermomechanical limits with optomechanical buckled-dome microcavities

G. J. Hornig, K. G. Scheuer, E. B. Dew, R. Zemp, and R. G. DeCorby[*]

[1] ECE Department, University of Alberta, 9211-116 St. NW, Edmonton, AB, Canada, T6G 1H9
*Corresponding author: rdecorby@ualberta.ca

**We describe the use of monolithic, buckled-dome cavities as ultrasound sensors. Patterned delamination within a compressively stressed thin film stack produces high-finesse plano-concave optical resonators with sealed and empty cavity regions. The buckled mirror also functions as a flexible membrane, highly responsive to changes in external pressure. Owing to their efficient opto-acousto-mechanical coupling, thermal-displacement-noise limited sensitivity is achieved at low optical interrogation powers and for modest optical ($Q \sim 10^3$) and mechanical ($Q \sim 10^2$) quality factors. We predict and verify broadband (up to ~ 5 MHz), air-coupled ultrasound detection with noise-equivalent pressure (NEP) as low as ~ 30-100 $\mu$Pa/Hz$^{1/2}$. This corresponds to an ultrasonic force sensitivity ~ $2 \times 10^{-13}$ N/Hz$^{1/2}$ and enables the detection of MHz-range signals propagated over distances as large as ~ 20 cm in air. In water, thermal-noise-limited sensitivity is demonstrated over a wide frequency range (up to ~ 30 MHz), with NEP as low as ~ 100-800 $\mu$Pa/Hz$^{1/2}$. These cavities exhibit a nearly omnidirectional response, while being ~ 3-4 orders of magnitude more sensitive than piezoelectric devices of similar size. Easily realized as large arrays and naturally suited to direct coupling by free-space beams or optical fibers, they offer significant practical advantages over competing optical devices, and thus could be of interest for several emerging applications in medical and industrial ultrasound imaging.**

## 1. BACKGROUND AND INTRODUCTION

High-frequency ultrasound signals in the ~1 – 300 MHz range are widely used for medical imaging and non-destructive testing (NDT) [1,2]. Piezoelectric sensors and transducers are the dominant commercial technology for these applications. However, their sensitivity scales inversely with size [3,4], which creates challenges for high-resolution imaging at high ultrasound frequencies [5]. Several emerging applications (such as photoacoustic tomography [5]) are driving a need for broadband, omnidirectional, and ultra-high sensitivity detectors. Air-coupled ultrasound [6] places even greater requirements on the transducer sensitivity, due to a large mismatch in acoustic impedance at air/solid interfaces and the rapid increase of attenuation with frequency for ultrasound signals in air [7]. Nevertheless, air-coupled ultrasound is gaining in interest [8,9] since it underpins many important applications such as gas sensing [10] and non-contact imaging in both medicine [11] and industrial NDT [9].

Optical detection of ultrasound can deliver high sensitivity in a small footprint and is currently the subject of an intensive research effort [1,3,6,12-17]. Most of this work employs resonant optical structures (e.g., ring resonators [12,14] or photonic crystal microcavities [15,16]) to enhance readout sensitivity. Some devices rely on photoelastic/elasto-optic effects, where acoustic waves modulate the refractive index and/or dimensions of an optical medium [3,13]. For example, polymer ring resonators have achieved bandwidths up to ~ 350 MHz and low noise-equivalent pressure (e.g., NEP ~ 6 mPa/Hz$^{1/2}$ [14]) in water. The Fabry-Perot (FP) etalon has also played a central role [17]; notably, Guggenheim et al. [13] embedded spherical-mirror FP etalons into thick (30-450 $\mu$m) polymer layers, reporting water-coupled NEP as low as ~ 1.6 mPa/Hz$^{1/2}$. These polymer-based devices (both ring and FP resonators) typically have a directional response at high MHz frequencies [18], although a fiber-end facet device in Ref. [13] exhibited a nearly omnidirectional response attributed to its small optical interrogation volume.

Recently, sensors based on elasto-optic effects in a silicon photonics platform have also been pursued. Shnaiderman et al. [15] reported 'point' sensors, with acoustic signals incident onto the end facets of a 1-dimensional array of silicon waveguides (with embedded FP cavities), achieving water-coupled NEP as low as ~ 9 mPa/Hz$^{1/2}$ over ~ 230 MHz bandwidth. Hazan et al. [16] employed a polymer-coated silicon waveguide resonator, reporting water-coupled NEP as low as ~ 2.2 mPa/Hz$^{1/2}$ and bandwidth as high as ~ 200 MHz.

Cavity optomechanical devices [19,20] are a newer class of sensors, which combine resonant optical structures, such as FP or waveguide ring resonators, with resonant mechanical structures, such as vibrating membranes or cantilevers. In these devices, the optical resonance is used to enhance the detection sensitivity while the mechanical resonance is exploited to enhance the response to force, pressure, etc. and to enable detection at the thermomechanical noise limit [21,22]. Also, an oscillating mechanical element (such as a membrane) provides a much more favorable acoustic impedance match to air [7,10], compared to the photo-elastic devices discussed above.

Impressive milestones in optomechanical ultrasound sensing have already been reported. Westerveld *et al.* [12] achieved NEP as low as ~ 1.3 mPa/Hz$^{1/2}$ (in water) over a frequency range ~ 3-30 MHz, and a nearly omnidirectional response, using a 15-20 μm diameter membrane to modulate the effective index of a silicon waveguide ring resonator. Using microdisk resonators, Basiri-Esfahani *et al.* [23] reported NEP ~ 0.008 - 10 mPa/Hz$^{1/2}$ for detection (in air) of ultrasound frequencies up to 1 MHz, although thermomechanical-noise-limited detection was achieved only at a few mechanical resonance frequencies in that case. More recently [24], a microtoroid sensor was reported to have air-coupled NEP ~ 0.046 - 10 mPa/Hz$^{1/2}$ in the 0.25 - 3.2 MHz frequency range, but again with thermomechanical-noise-limited sensitivity near its fundamental resonance only. The restricted range of low NEP in those devices is in part due to the relatively inefficient coupling between acoustical, mechanical, and optical modes. Moreover, these and most of the other silicon devices cited above involve relatively complex fabrication processes, require inefficient or inconvenient optical coupling via tapered nanofibers or grating couplers, and in some cases do not appear to be easily scalable as 2-D sensor arrays.

Here, building on earlier static pressure-sensing experiments [25], we describe cavity optomechanical sensors which deliver record sensitivities while also providing key practical advantages such as straightforward fabrication of sensor arrays and convenient and direct coupling by optical fibers and beams. We demonstrate near-omnidirectional ultrasound detection at MHz frequencies, with sensitivities near the fundamental limits set by thermal-acoustic noise in the coupling medium. NEPs as low as ~ 30 μPa/Hz$^{1/2}$ and ~ 150 μPa/Hz$^{1/2}$ in air and water, respectively, are predicted and corroborated by experimental results. We furthermore discuss feasible routes towards future improvement.

## 2. DEVICE DESCRIPTION AND OPERATING PRINCIPLES

The optical and thermo-mechanical properties of our buckled cavities have been detailed elsewhere [25-28]. Briefly, they are fabricated by embedding circular patterns of a thin low-adhesion layer between two Bragg mirrors and subsequently inducing delamination buckles to form over these regions driven by compressive stress in the upper Bragg mirror. The 'spontaneous-assembly' nature of the process results in a highly predictable and smooth morphology, and the cavities tend to exhibit the 'textbook' Laguerre-Gaussian modes expected for a half-symmetric (plano-concave) spherical mirror cavity. These stable cavity modes (see the Supplementary Information file for example) are not subject to deleterious effects such as 'walk-off' [17], and can be efficiently coupled to a fiber mode or a focused beam.

Since the upper buckled mirror is a dense, sputtered thin-film multilayer, it also acts as a barrier to gas/liquid diffusion. Accordingly, the buckled features contain a partially evacuated cavity region [29] which is hermetically sealed from the external environment, as depicted in Fig. 1. The curved (buckled) membrane mirror also functions as a high-quality mechanical resonator [26], and (as shown below) is extremely responsive to acoustic signals.

We recently demonstrated the utility of these cavities as static pressure sensors [25]. Here, our goal is to assess their potential for detection of dynamic pressure signals, such as those used in ultrasound applications. To that end, we modeled (see Supplementary Information) the buckled mirror as a simple harmonic oscillator [30]. This approach is supported by our previous work [26-28], which established that the buckled mirror exhibits a sparse collection of vibrational modes lying in the MHz frequency range.

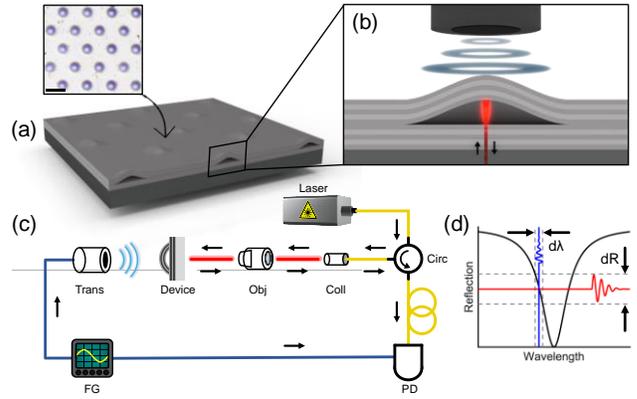

Fig. 1. (a) A depiction of an array of buckled dome etalons. Inset: Microscope image of an array of domes (scale bar: 100 μm). (b) Schematic illustration of the buckled-dome ultrasound sensor; an incident ultrasound pulse deflects the buckled mirror and results in a time-varying optical resonance, which is interrogated in reflection by a laser beam matched to the fundamental cavity mode (c) Schematic of the experimental setup. Trans: ultrasound transducer, Obj: objective lens, Coll: collimator, Circ: circulator, PD: photodetector, FG: function generator. (d) Schematic illustration of the tuned-to-slope readout mechanism, which maps external pressure signals to variations in the reflected optical power (see main text).

As mentioned briefly above, the detection limit of an optomechanical sensor is typically determined by laser shot noise and mechanical thermal noise (i.e., the natural vibrations of the mechanical element due to Brownian thermal noise) [19-22], rather than by electronic noise in the detection system. Moreover, for a sufficiently high-$Q$ optical cavity and/or a high optomechanical coupling coefficient, the impact of shot noise can also be negated (often at relatively low optical interrogation powers), resulting in detection at the fundamental limit set by thermal-displacement (i.e., thermomechanical [31]) noise [19].

Consider a dynamic pressure signal incident onto a buckled dome, as depicted schematically in Fig. 1(b). Assuming (*a priori*) operation in a thermal-displacement-noise-limited regime, and following the theoretical treatment in Ref. [25], a frequency-independent sensitivity is predicted from a single-harmonic oscillator model (see Supplementary Information for an expanded derivation and discussion) [31]:

$$NEP_{TD} = \frac{\sqrt{S_{xx}}}{\sigma_P} = \sqrt{\frac{2 k_B T \cdot k_{eff}}{\pi^3 \cdot a^4 \cdot f_0 \cdot Q}} \quad . \quad \textbf{(1)}$$

Here, $NEP_{TD}$ indicates NEP in the thermal-displacement-noise limit, $S_{xx}(f)$ is the thermal-vibrational displacement noise spectrum (in units of [m$^2$/Hz]) for a simple harmonic oscillator [30], $\sigma_P(f)$ is the pressure-induced motion of the buckled mirror (in units of [m/Pa]), $a$ is the base radius of the buckled dome, and $k_{eff}$, $f_0$, and $Q$ are the effective spring constant, vibrational resonance frequency, and quality factor of the mechanical oscillator, respectively. Furthermore, $f_0 = (k_{eff}/m_{eff})^{1/2}/2\pi$, where $m_{eff}$ is the effective mass of the vibrational mode [30] and $k_{eff} \sim (\pi \cdot a^2)/\sigma_P(0)$ [25].

We applied this simple theory to the two device sizes ('type *A*' with a 25 μm base radius and 'type *B*' with a 50 μm base radius) considered in our previous work [25], and for operation in both air and water (see Table 1). Representative (i.e., approximately average) values shown for $f_0$ and $Q$ were extracted from experimental noise spectra (see below), and the effective spring constants were estimated from static pressure experiments [25]. The mechanical $Q$ is mainly limited by viscous interactions with, and acoustic radiation into, the external coupling medium, and thus encapsulates the fundamental sensitivity limit set by acoustic noise [4,23]. For operation in water, the mechanical properties are modified relative to those in air [12,32]. First, the mechanical resonant frequency is reduced due to an increase in the effective mass. Second, the mechanical quality factor is reduced, primarily through acoustic radiation into the water medium. Experimentally observed values for $f_0$ and $Q$ in water are in good agreement with theoretical predictions (see Supplementary Information).

**Table 1. Predictions for Representative Cavities [a]**

| Device type | A (Ref. 33) | B (Ref. 34) |
|---|---|---|
| Base radius, $a$ | 25 [μm] | 50 [μm] |
| Mirror thickness, $h$ | 1.95 [μm] | 1.59 [μm] |
| Mirror mass, $m_B$ | 8.6 [ng] | 28 [ng] |
| Resonance freq. in air, $f_0$ | 10.9 [MHz] | 2.7 [MHz] |
| Resonance freq. in water, $f_0$ | 5.4 [MHz] | 0.8 [MHz] |
| Mechanical Q in air, $Q_A$ | 160 | 110 |
| Mechanical Q in water, $Q_W$ | 10 | 17 |
| DC pressure response, $\sigma_P$ | 0.04 [nm/kPa] | 1 [nm/kPa] |
| Spring constant, $K_{eff}$ | $4.9 \times 10^4$ [N/m] | $7.9 \times 10^3$ [N/m] |
| NEP$_{TD}$ in air, NEP$_{TD,A}$ | 135 [μPa/Hz$^{1/2}$] | 33 [μPa/Hz$^{1/2}$] |
| NEP$_{TD}$ in water, NEP$_{TD,W}$ | 780 [μPa/Hz$^{1/2}$] | 155 [μPa/Hz$^{1/2}$] |

[a] Static pressure sensing properties of device types A and B were detailed in Ref. [25].

The predicted $NEP_{TD}$ values are amongst the lowest reported for optical ultrasound sensors and are well corroborated by experimental results below. To achieve thermal-displacement-noise-limited sensitivity [19,21] generally requires some combination of high optical $Q$, high mechanical $Q$, high coupling between the pressure wave and the mechanical mode, and/or a high optomechanical coupling coefficient (i.e., a high value of $G = d\omega_c/dx$, where $\omega_c$ is the cavity resonance frequency and $x$ is the displacement of the mechanical part). Reliance on a high optical $Q$-factor [23,24] necessitates relatively sophisticated locking of the interrogation laser to the cavity resonance, while reliance on a high mechanical $Q$-factor can create challenges with respect to linearity and dynamic range [13].

Notably, our devices achieve displacement-noise-limited sensitivity over a wide frequency range, despite their modest optical and mechanical $Q$-factors, in large part due to their highly efficient coupling between a pressure wave, the mechanical modes, and the optical mode of interest. For example, the 'pressure participation ratio' and 'acousto-mechanical overlap factor', defined in Ref. [23], are both very close to the ideal value of unity for our devices. Moreover, $G$ is large in our devices, due to the direct correlation between mirror displacement and cavity resonance in a Fabry-Perot etalon.

## 3. EXPERIMENTAL SETUP AND CALIBRATION

Readout of the thermo-mechanical and acoustic response of the buckled dome cavities was accomplished using a tuned-to-slope technique [28], as depicted in Figs. 1(c) and (d). The device under study was interrogated by a single-mode-fiber-coupled tunable laser operating in the 1550 nm wavelength range (Santec TSL-710), passed through a fiber collimator and an objective lens, and focused onto the dome of interest through the double-side polished silicon substrate. Light reflected by the cavity was collected through the same optics and delivered via an optical circulator to a high-speed digitizing photodetector receiver (Resolved Instruments, DPD80) for analysis. The laser wavelength was tuned near the midway point of the Lorentzian lineshape associated with the fundamental cavity mode, such that dynamic changes in the cavity length (i.e., due to vibrations of the buckled mirror) were mapped to the received optical power (i.e., through 'dispersive' optomechanical coupling [12,19]).

Ultrasound signals were generated with a variety of standard piezo-based transducers (Olympus), centered at 2.25, 3.5, and 10 MHz and nominally designed for operation in water. For measurements in air, the transducers were driven by a commercial ultrasound pulse generator (Olympus 5800PR) with typical peak voltages of ~ 300 V. For the water-coupled measurements, it was necessary to drive the transducers with a much less energetic source, owing to the extreme sensitivity of our buckled dome devices. In that case, the transducers were driven by an arbitrary function generator (Rigol DG1022Z), with typical peak voltages on the order of 100 mV. Calibration of the ultrasound signals in the 1-20 MHz frequency range was accomplished in water by using a traceably calibrated hydrophone (Onda HNP-400). Device measurements in water were obtained for the same transducer spacing as during calibration, and the pressure spectral density (in units of [Pa/Hz$^{1/2}$]) was scaled by assuming a linear relationship between transducer drive voltage and pressure [12].

Calibration in air is more complicated due to the lack of readily available, calibrated microphones for operation in the MHz frequency range [6]. We followed the simple approach described in Ref. [24], wherein the pressure produced by a given transducer (and at a given drive voltage) is assumed to be scaled according to the medium impedance such that $P_A \sim P_W \cdot (Z_A/Z_W)$, where $P_W$ is the pressure measured in water, $P_A$ is the estimated pressure produced in the air at the transducer surface, and $Z_A$ and $Z_W$ are the acoustic impedances of air and water, respectively. Using $P_A$ and by accounting for the frequency-dependent acoustic loss in air at the transducer-device distance, the pressure incident onto the device ($P_D$) can then be estimated. The essential assumption in this calibration procedure is that the displacement of the transducer is the same in both air and water [23]. A more detailed description of the calibration procedure, for both water and air, is provided in the Supplementary Information.

## 4. AIR-COUPLED ULTRASOUND RESULTS

We first studied the detection of high-frequency ultrasound pulses delivered through an air medium. For the measurements described below, the 3.5 MHz transducer was driven with a high energy (100 μJ) electrical pulse. For the type A devices, the transducer-device spacing was set to ~ 5 mm (i.e., ~ 15 μs propagation delay), and the laser was adjusted near resonance and with ~ 10 μW of average power received by the photodetector. A typical time-domain trace, with the photodetector triggered by the pulse generator, is shown in Fig.4(a).

The corresponding frequency-domain content, shown as the red trace in Fig. 4(c), was obtained from the discrete Fourier transform (DFT) of the (300x averaged and bandpass filtered) pulse content lying between 14 and 16 μs [12]. This signal response is plotted alongside the background thermo-mechanical noise spectrum (the blue trace) extracted for the same cavity-laser detuning and laser power. This spectrum reveals the natural vibrational modes of the buckled mirror, with a fundamental resonance frequency at ~ 11 MHz and a second-order resonance near ~ 18 MHz. Also shown are the shot noise spectrum (the green trace), extracted from a signal trace with the laser detuned from the cavity resonance and with the same average optical power as above, and the spectrum of the photodetector dark noise (the black trace). Detailed information on the signal processing algorithms employed is provided in the Supplementary Information.

Analogous results are also shown for a type B device, but with the transducer-device spacing set to ~ 7 cm (i.e., ~ 200 μs propagation delay) in that case. The larger spacing provides preferential attenuation of the higher-frequency signal components, and thus reduced the 'ringing' caused by the overlap between the transducer's spectral content and the fundamental dome resonance at ~ 2.4 MHz. A typical time trace is shown in Fig. 2(b), and the corresponding frequency-domain content is shown in Fig.2(d), where a window from 200 to 207 µs was used for the DFT in this case.

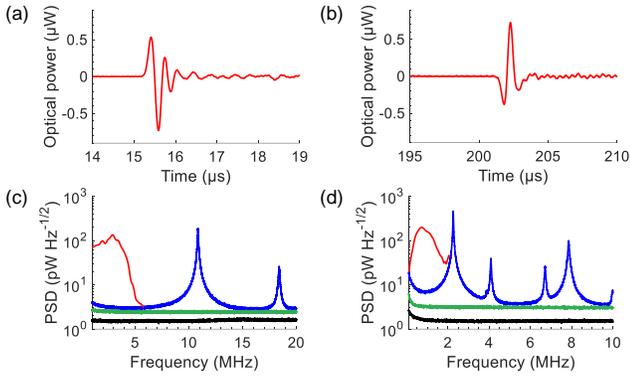

Fig. 2. Air coupled ultrasound pulses. (a,b) Time domain traces of the photodetector signal for device types *A* and *B*, respectively. Timing was triggered with respect to the electrical transducer pulse generated at $t = 0$. (c,d) The corresponding frequency-domain responses (red) for the above pulses shown alongside the background noise (blue,) the detuned optical shot noise (green,) and the PD dark noise (black) for device types *A* and *B*, respectively.

The results from Fig. 2, typical of measurements on many devices, immediately reveal a couple of interesting facts. First, with proper alignment to ensure good optical mode matching, a thermal-displacement-dominated noise floor [12,22] was observed over a wide frequency range, and for relatively low optical powers (<< 1 mW). Such behavior has not typically been easy to achieve in air [23,24]. Second, the devices enable high-SNR detection of ultrasound signals at frequencies well below their fundamental mechanical resonance. Moreover, since the frequency response is nearly flat in this regime, the received pulses are very similar to those recorded by the hydrophone. Non-periodic fluctuations beyond the duration of the main pulse, for example from 17 µs to 19 µs in Fig. 2(a), are consistent with reverberations in the underlying silicon substrate [12]. These reverberations cause signal distortion such as the jagged features in the frequency-domain trace of Fig. 2(c) and could be mitigated by proper substrate mounting. It is worth reiterating that the attenuation of ultrasound signals increases rapidly with frequency [7], effectively limiting the received pressure content to the frequency range below ~ 5 MHz for the transducer-device spacings studied. Further discussion is provided in the Supplementary Information file.

Figure 3 shows representative plots of the extracted sensitivity for these devices, obtained using the calibration procedures described above. Air-coupled NEP as low as ~ 100 and ~ 30 µPa/Hz$^{1/2}$ was estimated for device types *A* and *B*, respectively, in good agreement with the thermal-displacement-noise limited NEPs summarized in Table 1. The shaded bands in these plots represent an approximate range of uncertainty for $NEP_{TD}$ from Eq. (1), arising from the experimentally observed variations (over a large set of each type of device) in mechanical resonance frequency, quality factor, and effective spring constant [25]. The excellent agreement between theory and experiment, as well as the relatively flat sensitivity profile, indicates that the devices are in fact operating near the mechanical-thermal noise limit [12,31]. Residual variations in the NEP curves are partly attributable to hydrophone calibration uncertainty (particularly in the case of our air estimates) and the substrate acoustic reverberations mentioned above.

We were not able to perform calibrations in the frequency range below 1 MHz, due to limitations of our equipment. Nevertheless, the measured sensitivity is predicted to extend to very low acoustic frequencies, and this is supported by the dominance of the thermal-displacement noise floor down to frequencies in the few kHz region (not shown), below which the electronic noise begins to dominate. The utility of the devices at sub-MHz frequencies was qualitatively verified (see Supplementary Information), including experiments at human audible frequencies < 20 kHz, in which the devices were used to receive music signals and deliver them to an audio amplifier.

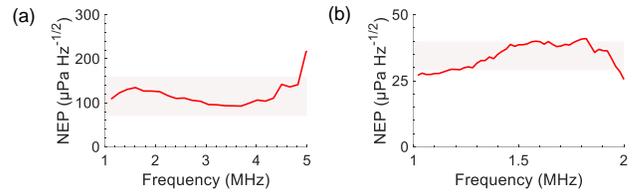

Fig. 3. Estimated NEP in air, for (a) Type *A* devices up to 5 MHz, and (b) Type *B* devices up to 2 MHz. The shaded areas indicate an uncertainty range of the theoretical sensitivity predicted by Eq. (1), accounting for observed variations in the mechanical properties of each family of devices (as explained in the main text).

The small size of these devices is expected to result in a nearly omni-directional response at MHz frequencies. To assess this, we mounted the 3.5 MHz transducer on a rotational stage and measured the device response at various angles, and for fixed transducer-device spacing and energy of the driving pulse. A typical result for a type *A* device is shown in Fig.4(a), where the 3.5 MHz transducer was driven with a 100 µJ pulse and the spacing was adjusted to ~ 1 cm (i.e., ~ 40 µs propagation delay) at each angle. Due to the large profile of the transducer, these measurements were restricted to the angular range below ~ 60 degrees. Nevertheless, an essentially non-directional response was verified in this range, and other observations (not shown) suggest that this response extends to near-glancing angles.

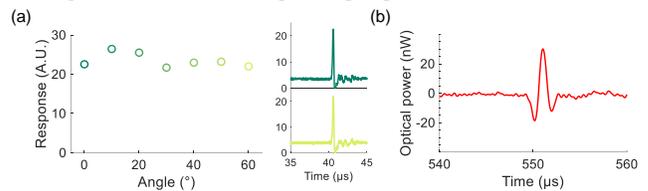

Fig. 4. (a) The directional dependence of the response of a typical type *A* device is plotted, for the case of the 3.5 MHz transducer driven by a 100 µJ pulse in air. Pulses detected at angles of 0 and 60 degrees (shown as insets) are nearly indistinguishable. (b) A pulse measured by a type *B* device for a transducer-device spacing of ~ 18 cm, and for the same transducer and driving conditions as in Fig. 2(b).

The unique combination of bandwidth, sensitivity, and omni-directionality described above might enable new applications for air-coupled ultrasound. As a further illustration, we note that these devices can detect MHz-range ultrasound pulses at distances in air which would traditionally be viewed as extreme. For example, Fig.4(b) shows a pulse received by a type *B* device, as generated by the 3.5 MHz transducer driven by a 100 µJ pulse, and for a transducer-device spacing of ~ 18 cm. Note that the SNR is still high, and the pulse character is very similar to that observed at shorter distances, such as shown in Fig. 2(b). This sensitivity could enable high-frequency (i.e., high resolution) air-coupled imaging and inspection, with relaxed requirements on the proximity between the sensor and the sample [9]. Moreover, these devices achieve NEPs in the MHz range comparable to the noise levels (a few µPa/Hz$^{1/2}$) associated with professional recording studios in the 0 - 20 kHz audio band [31]. This level of sensitivity might enable exotic applications such as photoacoustic monitoring of photosynthesis processes and detection of vibrations by individual cells [23].

## 5. WATER-COUPLED ULTRASOUND RESULTS

To accommodate experiments in water, the devices were mounted onto a fiberglass substrate and aligned to a pre-drilled hole to accommodate optical access. A plastic cylinder was glued to the same substrate to serve as a holding tank, and then filled with high purity deionized (DI) water. The ultrasound transducers were placed directly overtop the device chip, at a distance corresponding to a 50 μs propagation delay in water (∼ 7 cm). Ultrasound pulses were then measured and analyzed in the same way as described above for the air measurements, except that the transducers were driven by an arbitrary function generator to enable pulses of much lower energy.

Figure 5 shows a set of results for water-coupled ultrasound, analogous to the air-coupled results from Fig. 2. However, these results were obtained using the 10 MHz transducer driven by a 100 mV (peak) electrical pulse. Using hydrophone calibrations and linear scaling as described above, we estimated that the resultant ultrasound pulses have peak-to-peak pressure of ∼ 300 Pa. The time-domain signals shown in Figs. 5(a) and (b) were averaged across 300 received pulses. From the hydrophone measurements, the duration of the incident ultrasound pulse is on the order of a microsecond. The type $A$ devices reproduce this pulse characteristic reasonably well, although non-periodic oscillations persist beyond the ∼ 1 μs window, likely due to reverberations in the silicon substrate as discussed for the air case above. For the type $B$ devices, received pulses (even for larger distances or lower pulse energies than those discussed here) were clearly impacted by 'ringing' of the mechanical resonator, due in part to their higher response and the relatively high $Q$-factor of their fundamental resonance in water. With the transducers available, we were unable to sufficiently minimize these ringing effects, so that reliable estimation of the true device sensitivity was not possible and is left for future work. Thus, we restrict most of the discussion that follows to the less responsive, type $A$ devices.

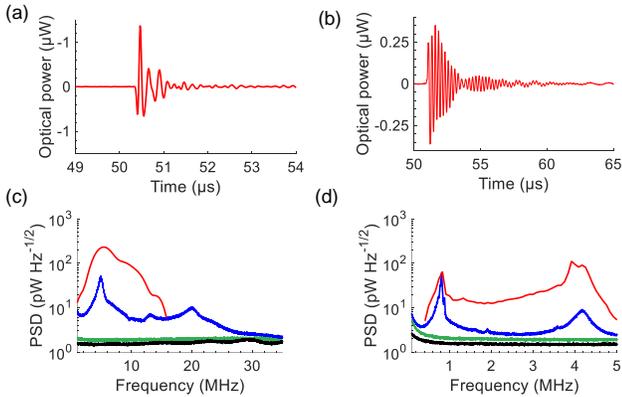

Fig. 5. Water-coupled ultrasound pulses. (a,b) Time domain traces of the (AC-coupled) optical signal for devices type $A$ and type $B$, respectively. (c,d) The corresponding frequency-domain responses for the above pulses (red) are shown alongside the background noise (blue), the detuned optical shot noise (green), and the PD dark noise (black) for device types $A$ and $B$, respectively.

As described for the air case above, the frequency-domain content of these pulses was analyzed by performing a DFT on a windowed portion (∼ 49-51 μs) of the time-domain traces. The resulting signal spectra (red traces) are plotted alongside the corresponding noise spectra in Figs. 5(c) and (d), for device types $A$ and $B$, respectively. The signal trace for the Type $A$ device clearly reflects the ∼1-16 MHz frequency content expected (from hydrophone calibrations) for the transducer used, with some resonant enhancement near 5 MHz. Consistent with the time-domain pulse, the signal spectrum for the type $B$ device is significantly impacted by the mechanical resonances.

The noise spectra in Fig. 5 reveal some notable properties of these devices. First, a large separation between the background noise and the shot noise floor is observed over a large frequency range, extending from ∼ 0 – 30 MHz for the Type $A$ devices. This suggests that broadband sensing at the thermal-displacement noise limit [12] should be possible. Second, while the reduction in mechanical resonance frequency and quality factor are consistent with the added mass and damping effects expected in water, we consistently observed an asymmetric and multiple-peaked character of the resonant modes for these devices in water. We attribute this to thermal noise 'crosstalk' or 'cross-coupling', likely in the form of acoustic radiation into the water medium, between neighboring devices in the closely spaced cavity arrays (see Fig. 1(a)). Cross-coupling between arrays of closely spaced and driven membranes is well-studied in ultrasound CMUT literature [35, 36]. However, crosstalk of thermal vibrational noise between neighboring non-driven devices has not to our knowledge been described previously and provides additional evidence for the extreme sensitivity and omni-directionality of the buckled domes as acoustic receivers. A more detailed discussion of this cross-coupling phenomenon is provided in the Supplementary Information.

The sensitivity of the type $A$ devices was extracted in a similar way as described for the air measurements, with a typical result shown in Fig. 6(a). Over much of the 0-15 MHz range considered, the extracted NEP is at or below the thermal-displacement-noise-limited value (∼ 780 μPa/Hz$^{1/2}$) listed in Table 1. However, near the fundamental resonance at ∼ 5 MHz there is a distinct spike in the NEP, which we attribute to the crosstalk discussed above. The rise in NEP near ∼13 MHz is similarly attributed to crosstalk mediated by higher-order mechanical modes. Aside from these crosstalk effects, the extracted and predicted NEP are in good agreement. In practice, thermal crosstalk noise could easily be reduced by isolating a single dome (e.g., by increased spacing or insertion of acoustic barriers). Furthermore, the noise spectrum in Fig. 5(c) suggests that thermal-displacement-limited sensitivity should extend up to ∼30 MHz.

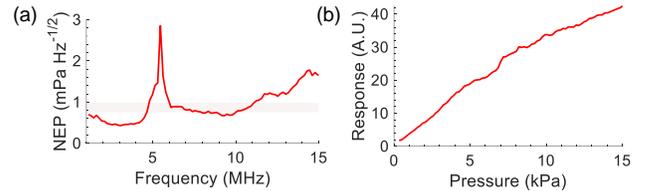

Fig. 6. (a) NEP for the type $A$ devices. As in Fig. 3, the shaded area indicates an uncertainty range of the theoretical sensitivity predicted by Eq. (1) given observed device variations. (b) Dynamic range study for the type $A$ devices, in which the peak response was tracked as the peak pressure produced by the 3.5 MHz transducer was varied. Transducer voltages were adjusted in 10 mV increments up to 3 V.

We also assessed the dynamic range of the type $A$ devices by varying the peak voltage of the electrical pulses applied to the 3.5 MHz transducer. The peak amplitude of the transduced signal as a function of peak pressure is plotted in Fig.6(b), revealing a linear relationship up to ∼ 5 kPa. The relatively low dynamic range is entirely a byproduct of the tuned-to-slope interrogation method [12], since pressure excursions larger than a few kPa result in the laser traversing a nonlinear portion of the cavity Lorentzian line (see Fig. 1(d)). This is consistent with the optical linewidth ($\Delta\lambda$ ∼ 0.8 nm) and wavelength-pressure response ($\sigma_\lambda = d\lambda/dP$ ∼ 0.1 nm/kPa) of our type $A$ devices [25]. Alternative, albeit more complex, interrogation methods could be employed to extend the linear dynamic range [12,37].

Finally, it is worth mentioning that many of the chips studied were repeatedly cycled between air and water immersion over several months and showed no degradation in their properties. Water immersion typically caused a slight shift in the optical resonance wavelengths, which would sometimes take a few hours to stabilize. This was attributed to stress changes in the sputtered films due to water infiltration and was reversible upon removal from water and with sufficient time for drying.

## 6. DISCUSSION AND CONCLUSIONS

We described a new class of optical ultrasound detectors which enable nearly omnidirectional ultrasound detection at frequencies in the MHz range, and sensitivities near the limits set by thermo-acoustic noise in an air or water coupling medium. These devices are highly practical, as their monolithic fabrication process produces arrays of high-quality, spherical mirror micro-resonators (with sealed cavity regions) which naturally support high-quality Gaussian beam modes and are thus easily and efficiently coupled to external optical fibers or lasers. It is also worth reiterating that these devices are quite robust, with the chips studied showing consistent performance over the course of several months and after repeated cycling between air and water coupling media. In some cases, chips were submerged in water for several weeks and showed no evidence of reduced performance.

Our results add to the growing body of evidence around the enabling properties of cavity optomechanical sensors [12,20-24]. Their resonant nature is sometimes cited as a drawback [6], since the frequency-dependent response can cause both dynamic range and signal fidelity issues in broadband (e.g., ultrasound) applications [13]. Nevertheless, for operation removed from the mechanical resonance, their extreme sensitivity can be exploited without incurring dynamic range issues [21]. Moreover, for narrowband applications aligned to the mechanical resonance, both extreme sensitivity and resonantly enhanced response can be exploited.

In water, our devices have sensitivity which compares favorably to electrical and optical sensors of similar size, as summarized in Table 2. The sensitivity is limited by thermal displacement noise, and we project that, with mitigation of inter-neighbor thermal crosstalk effects, the type $A$ devices can provide water-coupled NEP below 1 mPa/Hz$^{1/2}$ over the entire frequency range from DC to ~ 30 MHz.

**Table 2. Comparison of some small [a] water-coupled sensors**

| Device type | Bandwidth [MHz] | Size [a] [μm] | NEP [mPa/Hz$^{1/2}$] |
|---|---|---|---|
| Fiber hydrophone [38] | 25 | 25 [b] | 3200 |
| Needle hydrophone [39] | 100 | 40 | 1000 |
| Polymer micro-ring [14] | 350 | 60 | 5.6 |
| Capacitive (CMUT) [40] | 20 | 70 | 3.0 |
| Polymer/Si grating [16] | 200 | 30 | 2.2 |
| Plano-concave FP [13] | 40 | 10 [b] | 2.1 |
| Si membrane/ring [12] | 30 | 20 | 1.3 |
| This work, type $A$ | >15 | 50 [c] | 0.5 – 3.0 |

[a] 'Size' indicates the approximate (largest) lateral dimension of each device; we've defined 'small' here as size ~ 100 μm or less.
[b] Based on the optical mode diameter; the physical dimensions are much larger (>100 μm).
[c] Based on physical dimensions; however, this is likely an over-estimate given that the optical mode spot size is ~ 5 μm.

While their performance in water is compelling, it is perhaps their performance in air, as shown in Fig. 7, which is most remarkable. It should be noted that for the commercial 'akinetic' sensor [45] included in Fig. 7, we have used their stated NEP (~ 7 μPa/Hz$^{1/2}$) combined with the 2 mm × 2 mm area of their sensor 'frame' to estimate the force sensitivity. This estimate is likely somewhat generous, as discussed in detail in Ref. 23. At frequencies at or above 1 MHz, the ultrasonic force sensitivity of our devices (~ 2 × 10$^{-13}$ N/Hz$^{1/2}$) lies 1-2 orders of magnitude lower than that of previously demonstrated air-coupled sensors [23,45], and is moreover achieved over the entire ~ 0-5 MHz range. Such extreme force sensitivity could enable numerous applications in non-destructive testing, navigation, range-finding, gas leakage detection, and photoacoustic gas spectroscopy [23].

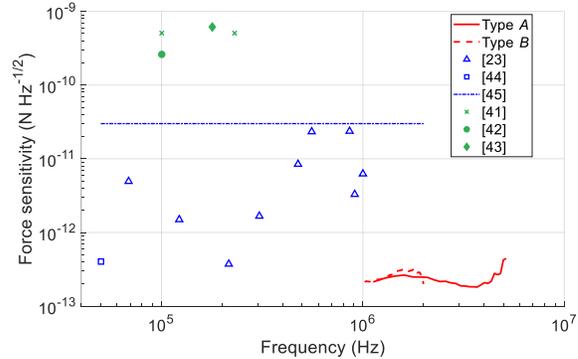

Fig. 7. Ultrasonic force sensitivity for comparative high frequency (> 50 kHz) air-coupled sensors. Results from this work are shown in red (solid and dashed for type $A$ and $B$, respectively,) while comparative results and their sources are indicated by the symbols shown in the legend. Blue or green symbols/lines indicate whether the comparator device is optical or electrical, respectively.

Finally, we note that there is significant scope for further improvements by engineering the mechanical properties of the buckled mirrors. The simple harmonic oscillator theory (see Eq. (1)), which is well corroborated by our experimental results, predicts that low $NEP_{TD}$ is favored by a low effective mass, a low spring constant, a large device radius, and a high mechanical quality factor, which are in some respects competing parameters. For example, the larger type $B$ devices achieve superior $NEP_{TD}$ primarily due to their lower effective spring constant, which results in a pressure response that is approximately an order of magnitude higher compared to the type $A$ devices but at the expense of a lower mechanical resonant frequency. Nevertheless, with further optimization of the devices, we project that significant reduction in $NEP_{TD}$ combined with an operational frequency range >> 30 MHz is feasible. Moreover, since the devices have already been realized as dense on-chip arrays, they offer potential for spatially resolved ultrasound imaging, and work is ongoing to address them individually through approaches such as using a 2-D fiber array or a focused scanning beam configuration [46]. We hope to explore these topics in future work.

**Funding Information.** Natural Sciences and Engineering Research Council of Canada (NSERC) (CREATE 495446-17); Alberta Innovates (SRP G2018000871).

**Acknowledgment**. We thank Tim Harrison, Lintong Bu, Afshin Ilkhechi, and Mahyar Ghavami for their assistance with experiments.

**Disclosures.** The authors declare no conflicts of interest.

**Data availability.** Data underlying the results presented in this paper are not publicly available at this time but may be obtained from the authors upon reasonable request.